\documentclass[12pt]{article}
\usepackage{graphicx}
\usepackage{amssymb}

\begin{document}
\title{Free Energy on a Cycle Graph and Trigonometric Deformation of
Heat Kernel Traces on Odd Spheres}
\author{
Nahomi Kan\\
{\footnotesize 
National Institute of Technology, Gifu College,
Motosu-shi, Gifu 501-0495, Japan}\\
{\small and}\\
Kiyoshi Shiraishi\\
{\footnotesize Graduate School of Sciences and Technology for Innovation,}
\\
{\footnotesize  Yamaguchi University, Yamaguchi-shi, Yamaguchi 753--8512,
Japan} }
\date{\today}
\maketitle

\abstract{We consider a possible `deformation' of the trace of the heat
kernel on odd dimensional spheres, motivated by the calculation of the
free energy of a scalar field on a discretized circle. By using an
expansion in terms of the modified Bessel functions, we obtain the values
of the free energies after a suitable regularization. 
}

\section{Introduction}
\label{intro}

Evaluation of functional determinants is an important task in quantum
field theories \cite{Dunne}.
To obtain the value of an infinite dimensional determinant,
we need appropriate regularization.

The heat kernel is one of the most powerful tools in mathematical
physics and spectral geometry 
\cite{Vassilevich,Gilkey,EORBZ,Elizalde,Kirsten,BCEMZ}.
We can regard the trace of the heat kernel (the heat kernel trace) as a
mathematical basis. Its advantage is that it can be computed at once
from the spectrum of the Laplacian operator defined in the theory.
The trace of the heat kernel is given by the sum of the Laplacian
eigenvalues exponentiated with the Schwinger `proper time' $t$.

The equivalence in computing the determinant by using the heat kernel
trace and by the zeta function technique
\cite{DC,Hawking} can be illustrated as follows.
The spectral zeta function $\zeta_S(s)$ for a differential operator on a
compact space with discrete eigenvalues $\lambda_l$ and degeneracies
(multiplicities)
$g_l$ is defined by the summation
\begin{equation}
\zeta_S(s)\equiv\sum_l
g_l\lambda_l^{-s}\,,
\end{equation}
and the log of the determinant is formally obtained by
\begin{equation}
-\zeta_S'(0)=-\left.\frac{\partial}{\partial
s}\zeta_S(s)\right|_{s=0}=\ln\prod_l\lambda_l^{g_l}\,.
\end{equation}

The spectral zeta function is written in an integral representation
\begin{equation}
\zeta_S(s)=\frac{\mu_R^{2s}}{\Gamma(s)}\int_0^\infty\frac{dt}{t^{1-s}}
\sum_l g_l\exp(-\lambda_l\, t)\,,
\end{equation}
where $\mu_R$ is a renormalization parameter and $\Gamma(z)$ is the gamma
function. In this expression,
\begin{equation}
\kappa(t)=\sum_l g_l\exp(-\lambda_l\, t)\,,
\end{equation}
is the trace of the heat kernel.

The one-loop free energy is then obtained by computing the log of the
determinant for the operator.
Once the spectral zeta function $\zeta_S(s)$ is known, the contribution of
the field under consideration to the free energy can be obtained.

It is known that the renormalization parameter $\mu_R$ only appear in even
dimensional space. Namely,
$\zeta_S(0)$ vanishes identically in odd dimensional space \cite{DT}.
Because $\Gamma(s)\sim\frac{1}{s}$ for small $s$,
the one-loop free energy of a scalar field $\phi$ on an odd dimensional
space can be evaluated as
\begin{equation}
F=-\ln Z=-\ln (\textrm{Det}\, D)^{-1/2}=-\frac{1}{2}\int_0^\infty\frac{dt}{t}
\,\textrm{Tr}\exp(-D\, t)\,.
\label{eq5}
\end{equation}
Here $Z$ is the partition function defined through the path integral
\begin{equation}
Z=\int {\cal D}\phi\, e^{-S}\,,
\end{equation}
where $S$ is the quadratic action for the scalar field described by
\begin{equation}
S=\frac{1}{2}\int d^dx\, \phi(x) D\phi(x)\,.
\end{equation}

Equation (\ref{eq5}) can also be understood using an `identity'%
\footnote{This can be confirmed by differentiating both sides of the
equation by $\lambda$.}
\begin{equation}
\ln \lambda\,\mbox{`}=\mbox{'}-\int_0^\infty\frac{dt}{t}e^{-\lambda t}\,,
\end{equation}
for each positive
eigenvalue
$\lambda$ of the operator
$D$.
This identity is correct up to an additional infinite constant,
which does not depend on $\lambda$ and
thus is irrelevant for physical quantities.
The infinity should be regularized by appropriate methods.

In this paper, we first review the derivation of the free energy of a
scalar field on a circle $S^1$ by using the heat kernel trace.
The Laplacian on $S^1$ has infinite discrete eigenvalues.

The discretization of space is discussed from time to time,
in various context, such as  approaches to quantum gravity
\cite{CaOrTh1,BaKeLo,Kerr1,Kerr2}, lattice (or coarse-graining)
simulations of field theory
\cite{Brower}, some exact supersymmetric models \cite{KS,MMO1,MMO2}, etc.
A naive discretization of
$S^1$ corresponds to defining a field theory on a cycle graph $C_N$
\cite{Wilson}, where $N$ is the number of vertices.
The eigenspectrum of the graph Laplacian of $C_N$ is known to be finite
and discrete.
In the present paper,
we show the regularization procedure of the free energy on $C_N$ by using
an expansion of the heat kernel trace in terms of the modified Bessel
functions. A key point of this expansion is the fact that the eigenvalues
of the graph Laplacian on $C_N$ can be written by the trigonometric
functions, such as $\propto 4\sin^2\frac{\pi p}{N}$ $(p=0,1,\dots N-1)$.

Motivated by this calculation, we come to the idea to deform the
eigenvalues as well as degeneracies in heat kernel traces on higher
dimensional space, by use of trigonometric functions. We demonstrate the
trigonometric deformation of the heat kernel traces and obtain
regularized free energies on odd dimensional spheres in the present paper.

Our trigonometric deformation is almost equivalent to $q$-deformation
\cite{KaCh}.
\footnote{As a mathematical term, the word `deformation' usually
represents for varying algebraic relations. In this paper, `deformation'
is used rather in a general sense.} 
The (symmetric)
$q$-number is defined by%
\footnote{Although the $q$-number is often denoted by $[p]_q$, the
simpler notation $[p]$ is adopted throughout the present paper.}
\begin{equation}
[p]=\frac{q^{p/2}-q^{-p/2}}{q^{1/2}-q^{-1/2}}\,.
\end{equation}
In the `classical' limit $q\rightarrow 1$, $[p]=p$ is recovered.
If $q$ is chosen as a primitive root of unity of degree $N$, i.e.,
$q=e^{i\frac{2\pi}{N}}$, it is found out to be
\begin{equation}
[p]=\frac{\sin\frac{\pi p}{N}}{\sin\frac{\pi}{N}}\,.
\end{equation}
There is, however, an ambiguity of the interpretation in extensions.
We give a comment on this ambiguity later in the present paper.

The outline of this paper is as follows.
In Section~\ref{fs1} we calculate the free energy of a scalar field on
$S^1$ by utilizing the heat trace method.
Section~\ref{fcn}, we consider the scalar field on $C_N$ and 
develop the regularization technique of the free energy.
In Section~\ref{defhk} we propose a trigonometric deformation of the heat
kernel traces of scalar fields on odd spheres. In Section~\ref{calcdef} we
perform an explicit calculation of the free energy of a scalar field on
$S^3$ and calculations of the free energies by the deformed heat kernel
traces with a suitable regularization. A comment on the ambiguity in
extensions is given in Section~\ref{qcom}. Finally, Section~\ref{conc}
offers some conclusions and directions for future research.

\section{Free energy on $S^1$}
\label{fs1}

We begin by the following Euclidean action for a massive scalar field
with mass $m$ on $S^1$:
\begin{equation}
S=\frac{1}{2}\int dx \left({\phi'}^2+m^2\phi^2\right)\,,
\label{SS}
\end{equation}
where we denote the derivative with respect to $x$  by the prime ($'$).
If the periodic condition $\phi(x)=\phi(x+2\pi a)$ is assumed,
i.e., $a$ is the radius of the circle,
the eigenvalue of the operator $D=-\partial_x^2+m^2$ is found to be
\begin{equation}
\lambda_l(m)=\frac{l^2}{a^2}+m^2\,,
\end{equation}
where $l$ is an integer ($l\in\mathbb{Z}$).

Now, we compute the free energy by the heat kernel method.
To this purpose, recall the relation of one-loop partition function with
traced heat kernel and we get the expression
\begin{equation}
F(m)=-\ln Z=-\frac{1}{2}\int_0^\infty\frac{dt}{t}\sum_{l\in
\mathbb{Z}}e^{-a^2\lambda_l(m)\,t}\,.
\end{equation}
Here, we note that we take $a^2\lambda_l(m)$ in stead of
$\lambda_l(m)$ in the exponential function because of a reparametrization
invariance on
$t$.
Then, the calculation is straightforward as
\begin{eqnarray}
F(m)&=&-\frac{1}{2}\int_0^\infty\frac{dt}{t}\sum_{l\in
\mathbb{Z}}e^{-(l^2+m^2a^2)t}\nonumber \\
&=&-\frac{\sqrt{\pi}}{2}\int_0^\infty\frac{dt}{t^{3/2}}
\sum_{l\in\mathbb{Z}}\exp\left(-m^2a^2 t-\frac{\pi^2l^2}{t}\right)
\nonumber \\
&=&-2\sum_{l=1}^\infty\sqrt{\frac{ma}{l}}K_{1/2}(2\pi
mal)-\frac{\sqrt{\pi}}{2}\int_0^\infty\frac{dt}{t^{3/2}}e^{-m^2a^2 t}
\nonumber \\
&=&-2\sum_{l=1}^\infty\sqrt{\frac{ma}{l}}\sqrt{\frac{\pi}{4\pi
mal}}e^{-2\pi
mal}-\frac{\sqrt{\pi}}{2}\Gamma\left(-\frac{1}{2}\right)ma
\nonumber \\
&=&-\sum_{l=1}^\infty\frac{e^{-2\pi mal}}{l}+\pi ma
=\ln(1-e^{-2\pi ma})+\ln e^{\pi ma}
\nonumber \\
&=&\ln(2\sinh\pi ma)\,.
\end{eqnarray}
Here we have used the Poisson summation formula
\begin{equation}
\sum_{l\in\mathbb{Z}}e^{-l^2t}=\sqrt{\frac{\pi}{t}}\sum_{l\in
\mathbb{Z}}e^{-\frac{\pi^2l^2}{t}}\,,
\end{equation}
and the integration formula
\begin{equation}
K_\nu(z)=\frac{1}{2}\left(\frac{z}{2}\right)^\nu
\int_0^\infty\exp\left(-t-\frac{z^2}{4t}\right)t^{-\nu-1}dt\,,
\end{equation}
\begin{equation}
K_{1/2}(z)=\sqrt{\frac{\pi}{2z}}e^{-z}\,,
\end{equation}
where $K_\nu(z)$ is the modified Bessel function of the second kind
(Macdonald function) \cite{GrRy}.

Incidentally, the regularized free energy on $S^1$ can also be directly
obtained but if we introduce the mass scale $M$ of Pauli-Villars
regularization
\cite{Monin}, and is found to be
\begin{equation}
F^{PV}=\frac{1}{2}\ln\prod_{l\in
\mathbb{Z}}\frac{\lambda_l(m)}{\lambda_l(M)}=\ln\left(\frac{m}{M}
\prod_{l=1}^\infty\frac{l^2+m^2a^2}{l^2+M^2a^2}\right)
=\ln\frac{2\sinh\pi ma}{2\sinh\pi Ma}\,,
\end{equation}
and this reads
\begin{equation}
F^{PV}=F(m)-F(M)\,,
\end{equation}
and thus, we have confirmed the previous result of the heat kernel
method. 

In the next section, we consider the discretization of the circle
and calculate the free energy on it.
We will see the continuum limit recovers the free energy on $S^1$.

\section{Free energy on a cycle graph} 
\label{fcn}

\begin{figure}[ht]
\centering
\includegraphics[height=3.5cm]
{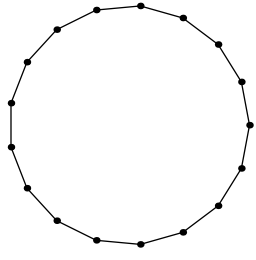}
\caption{%
$C_{17}$ as an example of a cycle graph.
}
\label{fig1}
\end{figure}

The cycle graph $C_N$ consists of $N$ vertices ($v\in V$) and $N$ edges
($e\in E$) connecting two adjacent vertices (see FIG.~\ref{fig1})
\cite{Wilson}.

We consider $N$ scalar degrees of freedom associated with $N$ vertices
and define the action as follows:
\begin{equation}
S=\frac{1}{2}\left(\frac{1}{a_0^2}\sum_{v,v'\in
V}\phi_v\Delta^G_{vv'}\phi_{v'}+
\mu^2\sum_{v\in V}\phi_v^2\right)\,,
\label{SC}
\end{equation}
where $\Delta^G$ is the graph Laplacian for $C_N$
\cite{Mohar1,Mohar2,Mohar3,Merris}. This is an $N\times N$ matrix
\begin{equation}
\Delta^G=\left(
\begin{array}{rrrrrr}
2 & -1 & 0 & \cdots & 0 & -1\\
-1 & 2 & -1 & \cdots & 0 & 0\\
0 &-1 & 2 & \cdots & 0 & 0\\
\vdots & \vdots & \vdots & \ddots & \vdots & \vdots\\
0 & 0 & 0 & \cdots  & 2 & -1 \\
-1 & 0 & 0 & \cdots  &-1 & 2 
\end{array}
\right)\,.
\end{equation}
For later convenience, we define the `mass' $\mu$ as
\begin{equation}
\mu\equiv\frac{2}{a_0}\sinh\frac{ma_0}{2}\,.
\label{choice}
\end{equation}
Note that the action (\ref{SC}) becomes equivalent to the action
(\ref{SS}) in the limit of $N\rightarrow\infty$ and $a_0\rightarrow 0$
while $a=Na_0/(2\pi)$ is left constant.

Then, the eigenvalue of the operator $\Delta^G+\mu^2$
is expressed by
\begin{equation}
\bar{\lambda}_p(m)=\frac{4}{a_0^2}\sin^2\frac{\pi p}{N}+
\frac{4}{a_0^2}\sinh^2\frac{\pi ma}{N}\,,
\end{equation}
where the integer $p$ runs over $0\le p\le N-1$.

Defining $\theta_p\equiv \frac{2\pi p}{N}$,
the free energy on $C_N$ can be expressed as
\begin{eqnarray}
F_{C_N}&=&-\frac{1}{2}\int_0^\infty\frac{dt}{t}\sum_{p=0}^{N-1}
e^{-a_0^2\bar{\lambda}_p
t}=-\frac{1}{2}\int_0^\infty\frac{dt}{t}\sum_{p=0}^{N-1}
e^{-(2-2\cos\theta_p+4\sinh^2\frac{\pi ma}{N}) t}
\nonumber \\
&=&-\frac{1}{2}\int_0^\infty\frac{dt}{t}\sum_{p=0}^{N-1}
\sum_{l\in\mathbb{Z}}\cos l\theta_p
I_l(2t)e^{-2(\cosh\frac{2\pi ma}{N}) t}
\nonumber \\
&=&-\frac{1}{2}\int_0^\infty\frac{dt}{t}
\sum_{l\in\mathbb{Z}}\sum_{p=0}^{N-1}e^{il\theta_p}
I_l(2t)e^{-2(\cosh\frac{2\pi ma}{N}) t}
\nonumber \\
&=&-\frac{N}{2}\int_0^\infty\frac{dt}{t}
\sum_{q\in\mathbb{Z}}
I_{Nq}(2t)e^{-2(\cosh\frac{2\pi ma}{N}) t}
\nonumber \\
&=&-N\sum_{q=1}^\infty\int_0^\infty\frac{dt}{t}
I_{Nq}(2t)e^{-2(\cosh\frac{2\pi ma}{N})
t}-\frac{N}{2}\int_0^\infty\frac{dt}{t} I_0(2t)e^{-2(\cosh\frac{2\pi
ma}{N}) t}\,,
\label{Cs}
\end{eqnarray}
where we have used the formula including the modified Bessel
function of the first kind $I_\nu(z)$ \cite{PrBrMa2}%
\footnote{This expansion formula has been utilized for calculating vacuum
energies in models of dimensional deconstruction
\cite{KaSaSh,ShSaKa,Ka1,Ka2,Ka3}.}
\begin{equation}
e^{z\cos\theta}=\sum_{l\in\mathbb{Z}}\cos l\theta I_l(z)\,,
\end{equation}
and $I_l(z)=I_{-l}(z)$ for $l\in\mathbb{Z}$.

By using the integration formula \cite{PrBrMa2}
\begin{equation}
\int_0^\infty\frac{dt}{t}
I_\nu(2t)e^{-2(\cosh x) t}=\frac{e^{-\nu x}}{\nu}\,,
\label{if}
\end{equation}
we can evaluate the convergent sum in the first term in the last
line of (\ref{Cs}) as
\begin{equation}
-N\sum_{q=1}^\infty\int_0^\infty\frac{dt}{t}
I_{Nq}(2t)e^{-2(\cosh\frac{2\pi
ma}{N}) t}=-\sum_{q=1}^\infty\frac{(e^{-2\pi ma})^q}{q}
=\ln(1-e^{-2\pi ma})\,.
\end{equation}

Since the second term  in the last line of (\ref{Cs}) diverges,
we should regularize this contribution.
We interpret this presentation as
\begin{equation}
-\frac{N}{2}\lim_{\epsilon\rightarrow 0}\int_0^\infty\frac{dt}{t}
I_\epsilon(2t)e^{-2(\cosh\frac{2\pi ma}{N}) t}\,.
\end{equation}
Then, this reads
\begin{equation}
-\frac{N}{2}\lim_{\epsilon\rightarrow 0}\frac{e^{-\frac{2\pi
ma}{N}\epsilon}}{\epsilon}
=-\frac{N}{2}\lim_{\epsilon\rightarrow
0}\frac{1}{\epsilon}\left(1-\frac{2\pi ma}{N}\epsilon\right)
=-\frac{N}{2}\lim_{\epsilon\rightarrow
0}\frac{1}{\epsilon}+\pi ma
\,.
\label{div}
\end{equation}

Dropping the first divergent term in (\ref{div}),
we can obtain the regularized free energy on $C_N$
\begin{equation}
F_{C_N}=\ln(1-e^{-2\pi ma})+\pi ma=\ln(2\sinh\pi ma)\,.
\end{equation}
Now, this expression is the same as that of free energy on $S^1$.
Our choice (\ref{choice}) of $\mu$ implies the $N$-independent free
energy.

\section{Heat kernel traces on odd spheres and their trigonometric
analogue}
\label{defhk}

In this section, first we compute the trace of the heat kernel of
scalar field theories on odd dimensional spheres. Next we propose the
trigonometric analogue of the heat kernel traces.

\subsection{eigenvalues and degeneracies on odd spheres}

For $d\ge 2$, let $S^d$ be the $d$-dimensional sphere with the standard
metric $g_{\mu\nu}$ $(\mu,\nu=1,\dots,d)$, and let $\Delta$ be the
Laplace-Beltrami operator on the space of smooth functions.

Consider the action for a scalar field in the
continuum on
$S^d$ given by
\begin{equation}
S=\frac{1}{2}\int d^dx
\sqrt{g}\left[\nabla_\mu\phi\nabla^\mu\phi+(m^2+\xi
{\cal R})\phi^2\right]\,,
\end{equation}
where $g=\det g_{\mu\nu}$.
The Laplacian acting on the scalar field is 
\begin{equation}
\Delta=\nabla^\mu\nabla_\mu=
\frac{1}{\sqrt{g}}\partial_\mu\sqrt{g}g^{\mu\nu}\partial_\nu\,.
\end{equation}
The scalar curvature ${\cal R}$ takes the value
\begin{equation}
{\cal R}=\frac{d(d-1)}{a^2}\,,
\end{equation}
on $S^d$ with the radius $a$.
The Laplacian has a discrete and infinite spectrum.
The eigenvalue of the operator $-\Delta+m^2+\xi{\cal R}$ is known to be
\cite{ErMaObTr}
\begin{eqnarray}
a^2\lambda_l&=&l(l+d-1)+\xi d(d-1)+m^2a^2
\nonumber \\
&=&\left(l+\frac{d-1}{2}\right)^2
-\frac{(d-1)^2}{4}+\xi d(d-1)+m^2a^2\,,
\end{eqnarray}
where $l$ is a non-negative integer, i.e., $l=0, 1, 2, \dots$.

We consider two specific cases of
the coupling $\xi$ to the scalar curvature ${\cal R}$.
One is the case with the conformal coupling
\begin{equation}
\xi=\xi_d=\frac{d-2}{4(d-1)}\,.
\end{equation}
In this case, the eigenvalue becomes
\begin{eqnarray}
a^2\lambda_l&=&\left(l+\frac{d-2}{2}\right)\left(l+\frac{d}{2}\right)+m^2a^2
\nonumber \\
&=&\left(l+\frac{d-1}{2}\right)^2
-\frac{1}{4}+m^2a^2\,.
\end{eqnarray}
It is known that the scalar field theory with the coupling $\xi_d$ has
conformal invariance at zero mass ($m=0$).

Another case is with the `pseudo-conformal coupling' defined as
\begin{equation}
\xi=\xi_{d+1}=\frac{d-1}{4d}\,.
\end{equation}
In this case, the eigenvalue can be written as 
\begin{equation}
a^2\lambda_l=\left(l+\frac{d-1}{2}\right)^2+m^2a^2\,.
\end{equation}

The corresponding degeneracy (multiplicity) of the $l$ th eigenvalue is
independent of $\xi$ and given by
\cite{ErMaObTr}
\begin{equation}
g_l=\frac{2\left(l+\frac{d-1}{2}\right)}{(d-1)!}\prod_{n=1}^{d-2}(l+n)\,.
\label{gl}
\end{equation}
Especially, for any $d$-sphere, we find
\begin{equation}
g_0=1\,.
\end{equation}
For instance, the expression (\ref{gl}) reads
\begin{equation}
g_l=(l+1)^2\,,\qquad\mbox{for $d=3$}\,,
\end{equation}
and
\begin{equation}
g_l=\frac{1}{12}(l+2)^2(l+3)(l+1)=\frac{1}{12}(l+2)^2[(l+2)^2-1]\,,
\qquad\mbox{for $d=5$}\,.
\end{equation}

The degeneracy for odd spheres can be rewritten as \cite{CaWe}
\begin{equation}
g_l=\frac{2\left(l+v\right)^2}{(2v)!}
\prod_{n=1}^{v-1}\left[\left(l+v\right)^2-n^2\right]\,,
\end{equation}
where we set
\begin{equation}
v\equiv\frac{d-1}{2}\,.
\end{equation}
Note that $v$ is a positive integer for odd $d$.

Then, the heat kernel trace of massless conformal scalar field on $S^d$
($d$ odd) can be expressed as
\begin{eqnarray}
\kappa_v(t)&=&\sum_{l=0}^\infty g_l \exp\left\{-a^2\lambda_l t\right\}
\nonumber \\
&=&\sum_{l=0}^\infty \frac{2\left(l+v\right)^2}{(2v)!}
\prod_{n=1}^{v-1}\left[\left(l+v\right)^2-n^2\right]
\exp\left\{-\left[\left(l+v\right)^2
-\frac{1}{4}\right]t\right\}
\nonumber \\
&=&\sum_{l=v}^\infty \frac{2l^2}{(2v)!}
\prod_{n=1}^{v-1}\left(l^2-n^2\right)
\exp\left[-\left(l^2
-\frac{1}{4}\right)t\right]
\nonumber \\
&=&\frac{1}{2}\sum_{l\in\mathbb{Z}}\frac{2l^2}{(2v)!}
\prod_{n=1}^{v-1}\left(l^2-n^2\right)
\exp\left[-\left(l^2
-\frac{1}{4}\right)t\right]\,.
\label{smf}
\end{eqnarray}
The last equality is due to
$l^2\prod_{n=1}^{v-1}\left(l^2-n^2\right)=0$ for
$l=0,1,\dots,v-1$.

In the next subsection, we consider a trigonometric analogue of the
heat kernel traces on odd spheres.

\subsection{trigonometric deformation}

We first consider a trigonometric analogue of eigenvalues of operators.
Following the naive correspondence between the cases with $S^1$ and $C_N$,
we replace eigenvalues for each case as follows.

For the case with conformal coupling, we replace the eigenspectrum as
\begin{eqnarray}
a_0^2\bar{\lambda}_p&=&4\sin\frac{\pi}{N}\left(p+\frac{d-2}{2}\right)
\sin\frac{\pi}{N}\left(p+\frac{d}{2}\right)+\mu^2a_0^2
\nonumber \\
&=&2\left[\cos\frac{\pi}{N}-\cos\frac{2\pi}{N}\left(p+\frac{d-1}{2}\right)
\right]+\mu^2a_0^2\,,
\end{eqnarray}
where $\mu$ and $a_0$ is constants corresponding to the mass of the
scalar field and the radius of $S^d$, respectively. Here $N$ is an
arbitrary integer, which is considered here to be sufficiently large. For
the case with pseudo-conformal coupling, we set
\begin{eqnarray}
a_0^2\bar{\lambda}_p&=&4\sin^2\frac{\pi}{N}\left(p+\frac{d-1}{2}\right)+\mu^2a_0^2
\nonumber \\
&=&2\left[1-\cos\frac{2\pi}{N}\left(p+\frac{d-1}{2}\right)
\right]+\mu^2a_0^2\,.
\end{eqnarray}
In both two cases, $p$ runs from $0$ to $N-1$.
Incidentally, these replacement is equivalent to considering $q$-analogue
of $x$,
$[x]=\frac{\sin\frac{\pi x}{N}}{\sin\frac{\pi}{N}}$.%
\footnote{We have not to care about the normalization due to
$\sin\frac{\pi}{N}$ etc., because they can be absorbed into the
redefinition of the parameter $t$.}

We also propose that the degeneracy is also deformed.
In order to normalize the degeneracy which satisfies $g_0=1$,
we consider the following form: 
\begin{equation}
\bar{g}_p\equiv\frac{\check{g}_p}{\check{g}_0}\,,
\end{equation}
where
\begin{eqnarray}
\check{g}_p&=&\sin\frac{\pi}{N}\left(p+v\right)
\prod_{n=-(v-1)}^{v-1}\sin\frac{\pi}{N}
\left(p+v+n\right)
\nonumber \\
&=&\frac{1}{2^{v}}\left[1-\cos\frac{2\pi}{N}\left(p+v\right)\right]
\prod_{n=1}^{v-1}\left[\cos\frac{2\pi n}{N}-\cos\frac{2\pi}{N}
\left(p+v\right)\right]
\nonumber \\
&=&\frac{1}{2^{v}}
\prod_{n=0}^{v-1}\left[\cos\frac{2\pi n}{N}-\cos\frac{2\pi}{N}
\left(p+v\right)\right]\,.
\end{eqnarray}
Note that $g_p>0$ for $p=0,\dots,N-1$.

Now, the trigonometric deformation of the heat kernel trace of the
massless conformal coupling scalar field on $S^d$ can be written as
\begin{eqnarray}
\bar{\kappa}_v(t)&=&\frac{1}{2}\sum_{p=0}^{N-1}
\prod_{n=0}^{v-1}\frac{\cos\frac{2\pi n}{N}-\cos\frac{2\pi}{N}
\left(p+v\right)}{\cos\frac{2\pi n}{N}-\cos\frac{2\pi v}{N}}
\exp\left\{-2\left[\cos\frac{\pi}{N}-\cos\frac{2\pi}{N}\left(p+v\right)
\right]t\right\}
\nonumber \\
&=&\frac{1}{2}\sum_{p=0}^{N-1}
\prod_{n=0}^{v-1}\frac{\cos\theta_n
-\cos\theta_p}{\cos\theta_n-\cos\theta_{v}}
\exp\left[-2\left(\cos\frac{\pi}{N}-
\cos\theta_p
\right)t\right]\,,
\label{defk}
\end{eqnarray}
where $\theta_k\equiv\frac{2\pi}{N}k$. Similarly the case with
pseudo-conformal coupling is treated. In the last equality in
(\ref{defk}), permutation invariance of the trace (sum over $N$ terms)
under
$p\rightarrow p+(any~integer)$ has been used. Note that the factor $1/2$
in front of the summation is needed because of
$\cos\theta_p=\cos\theta_{N-p}$.

One can check that, in the limit of $N\rightarrow\infty$,
the deformed heat kernel trace becomes the one on $S^d$, up to the
redefinition of the parameter $t$.

\section{Free energies on odd spheres and their trigonometric
deformations}
\label{calcdef}

In this section, we first explicitly demonstrate the calculation of the
free energy on $S^3$ by using the heat kernel traces, of standard and
deformed. Next, we give the regularized `deformed' free energies of
general odd dimensions.

\subsection{free energy on $S^3$} 

First, we show the evaluation of the free energy on $S^3$ via the heat
kernel trace method.
Using the Poisson summation formula, we have
\begin{eqnarray}
\sum_{l\in\mathbb{Z}}l^2 e^{-l^2 t}&=&-\frac{\partial}{\partial t}
\sum_{l\in\mathbb{Z}}e^{-l^2 t}
=-\frac{\partial}{\partial t}\left(\sqrt{\frac{\pi}{t}}
\sum_{l\in\mathbb{Z}}e^{-\frac{\pi^2l^2}{t}}\right)
\nonumber \\
&=&\frac{\sqrt{\pi}}{2t^{3/2}}
\sum_{l\in\mathbb{Z}}e^{-\frac{\pi^2l^2}{t}}
-\frac{\pi^{5/2}}{t^{5/2}}
\sum_{l\in\mathbb{Z}}l^2 e^{-\frac{\pi^2l^2}{t}}\,.
\end{eqnarray}

For the conformal coupling case, if we define
$y^2\equiv \mu^2a^2-\frac{1}{4}$, we can proceed to calculate
the free energy
with the above relation, as
\begin{eqnarray}
F_3&=&-\frac{1}{2}\int_0^\infty\frac{dt}{t}\frac{1}{2}
\sum_{l\in\mathbb{Z}}l^2e^{-l^2 t-y^2t}
\nonumber \\
&=&-\frac{\sqrt{\pi}}{8}\int_0^\infty\frac{dt}{t^{5/2}}
\sum_{l\in\mathbb{Z}}e^{-y^2t-\frac{\pi^2l^2}{t}}
+\frac{\pi^{5/2}}{4}\int_0^\infty\frac{dt}{t^{7/2}}
\sum_{l\in\mathbb{Z}}l^2 e^{-y^2t-\frac{\pi^2l^2}{t}}
\nonumber \\
&=&-\frac{\sqrt{\pi}}{2}\sum_{l=0}^\infty
\left(\frac{y}{\pi l}\right)^{3/2}
K_{3/2}(2\pi yl)
-\frac{\sqrt{\pi}}{8}\int_0^\infty\frac{dt}{t^{5/2}}
e^{-y^2t}
\nonumber \\
& &+\pi^{5/2}\sum_{l=0}^\infty l^2
\left(\frac{y}{\pi l}\right)^{5/2}
K_{5/2}(2\pi yl)\nonumber \\
\nonumber \\
&=&-\frac{\sqrt{\pi}}{2}\sum_{l=0}^\infty
\left(\frac{y}{\pi l}\right)^{3/2}
\frac{1}{\sqrt{4yl}}\left(1+\frac{1}{2\pi yl}\right)e^{-2\pi yl}
-\frac{\sqrt{\pi}}{8}\Gamma\left(-\frac{3}{2}\right) y^3
\nonumber \\
& &+\pi^{5/2}\sum_{l=0}^\infty l^2
\left(\frac{y}{\pi l}\right)^{5/2}
\frac{l^2}{\sqrt{4yl}}\left(1+\frac{3}{2\pi yl}+\frac{3}{4\pi^2
y^2l^2}\right)e^{-2\pi yl}
\nonumber \\
&=&\sum_{l=1}^\infty\left(
\frac{y^2}{2l}+\frac{y}{2\pi l^2}+\frac{1}{4\pi^2l^3}\right)e^{-2\pi yl}
-\frac{\pi}{6}y^3\,.
\label{s3}
\end{eqnarray}

Note that
\begin{equation}
K_{3/2}(z)=\sqrt{\frac{\pi}{2z}}\left(1+\frac{1}{z}\right)e^{-z}\,,
\end{equation}
and
\begin{equation}
K_{5/2}(z)=\sqrt{\frac{\pi}{2z}}\left(1+\frac{3}{z}
+\frac{3}{z^2}\right)e^{-z}\,.
\end{equation}

For a massless conformal scalar field, we have to set
$y\rightarrow\frac{i}{2}$. Then, we simply have
\begin{eqnarray}
F_3&=&\sum_{l=1}^\infty\left(
-\frac{1}{8l}+\frac{i}{4\pi l^2}+\frac{1}{4\pi^2l^3}\right)(-1)^{l}
+\frac{i\pi}{48}
\nonumber \\
&=&\left(
\frac{\ln
2}{8}-\frac{i}{4\pi}\eta(2)-\frac{1}{4\pi^2}\eta(3)\right)
+\frac{i\pi}{48}\nonumber \\
&=&
\frac{\ln
2}{8}-\frac{i}{8\pi}\zeta(2)-\frac{3}{16\pi^2}\zeta(3)
+\frac{i\pi}{48}
\nonumber \\
&=&
\frac{\ln
2}{8}-\frac{3\zeta(3)}{16\pi^2}\,.
\end{eqnarray}

Note that
\begin{equation}
\sum_{l=1}^\infty\frac{(-1)^{l-1}}{l}=\ln 2\,,\quad
\sum_{l=1}^\infty\frac{(-1)^{l-1}}{l^z}=\frac{2^{z-1}-1}{2^{z-1}}\zeta(z)
\equiv\eta(z)\quad (z\ge 2)\,,
\end{equation}
where $\eta(z)$ is known as the Dirichlet eta function
and $\zeta(z)$ is the Riemann zeta function. In particular,
it is known that $\zeta(2)=\frac{\pi^2}{6}$.

The free energy of a massless pseudo-conformal scalar field on $S^3$
is obtained when we set $y=0$ in (\ref{s3}). Then, we find
\begin{equation}
F_{3pc}=\frac{\zeta(3)}{4\pi^2}\,.
\end{equation}
Another derivation of free energies of massless pseudo-conformal scalar
fields on odd sphere is exhibited in Appendix A.

\subsection{the deformation of free energies} 

Next, we calculate the free energy by using the trigonometrically deformed
heat kernel trace. We have already seen the expansion formula
\begin{equation}
\sum_{p=0}^{N-1}e^{2t\cos\theta_p}=N\sum_{q\in\mathbb{Z}}
I_{Nq}(2t)\,,
\end{equation}
where $\theta_p=\frac{2\pi p}{N}$.
By differentiating both side of the equation, we get
\begin{eqnarray}
\sum_{p=0}^{N-1}\cos\theta_p\, e^{2t\cos\theta_p}&=&
\frac{\partial}{\partial
(2t)}\sum_{p=0}^{N-1}e^{2t\cos\theta_p}
=N\sum_{q\in\mathbb{Z}}
I'_{Nq}(2t)
\nonumber \\
&=&\frac{N}{2}\sum_{q\in\mathbb{Z}}
\left[I_{Nq-1}(2t)+I_{Nq+1}(2t)\right]\,,
\end{eqnarray}
where we have used the recurrence relation
\begin{equation}
I'_\nu(z)=\frac{1}{2}\left[
I_{\nu-1}(z)+I_{\nu+1}(z)\right]\,,
\end{equation}
and $I'_\nu(z)$ denotes the first derivative of $I_\nu(z)$.
Then, we have the expression by an infinite series, which appears in the
calculation of the free energy:
\begin{eqnarray}
& &\int_0^\infty\frac{dt}{t}\sum_{p=0}^{N-1}(1-\cos\theta_p)\,
e^{2t\cos\theta_p-2t\cosh x}
\nonumber \\
&=&N
\int_0^\infty\frac{dt}{t}\sum_{q\in\mathbb{Z}}\left\{I_{Nq}(2t)-
\frac{1}{2}
\left[I_{Nq-1}(2t)+I_{Nq+1}(2t)\right]\right\} e^{-2t\cosh x}\,.
\end{eqnarray}
The integration of each term over $t$ can be performed by use of the
formula (\ref{if}). Thus we find
\begin{eqnarray}
& &\int_0^\infty\frac{dt}{t}\left\{I_{\nu}(2t)-
\frac{1}{2}
\left[I_{\nu-1}(2t)+I_{\nu+1}(2t)\right]\right\} e^{-2t\cosh x}
\nonumber \\
&=&e^{-\nu x}\left[\frac{1}{\nu}-\frac{1}{2}\left(
\frac{e^x}{\nu-1}+\frac{e^{-x}}{\nu+1}\right)\right]
\nonumber \\
&=&e^{-\nu x}\left(\frac{1}{\nu}-
\frac{\nu \cosh x}{\nu^2-1}-\frac{\sinh x}{\nu^2-1}\right)\,,
\end{eqnarray}
and then, we obtain
\begin{eqnarray}
& &\int_0^\infty\frac{dt}{t}\sum_{p=0}^{N-1}(1-\cos\theta_p)\,
e^{2t\cos\theta_p-2t\cosh x}
\nonumber \\
&=&2N\sum_{q=1}^\infty
e^{-Nq x}\left(\frac{1}{Nq}-
\frac{Nq \cosh x}{N^2q^2-1}-\frac{\sinh
x}{N^2q^2-1}\right)
\nonumber \\
& &+N\lim_{\epsilon\rightarrow 0}e^{-\epsilon x}\left(\frac{1}{\epsilon}-
\frac{\epsilon \cosh x}{\epsilon^2-1}-\frac{\sinh x}{\epsilon^2-1}\right)
\nonumber \\
&=&2\sum_{q=1}^\infty
e^{-Nq x}\left(\frac{1}{q}-
\frac{q \cosh x}{q^2-1/N^2}-\frac{1}{N}\frac{\sinh
x}{q^2-1/N^2}\right)
\nonumber \\
& &-Nx+N\sinh x+N\lim_{\epsilon\rightarrow 0}
\frac{1}{\epsilon}\,,
\end{eqnarray}
where we have replaced only the term of $q=0$ by the expression with
$q=\epsilon\rightarrow 0$.

If we set $x\rightarrow \frac{i\pi}{N}$ in the above result,
\begin{eqnarray}
& &\int_0^\infty\frac{dt}{t}\sum_{p=0}^{N-1}(1-\cos\theta_p)\,
e^{2t\cos\theta_p-2t\cos\frac{\pi}{N}}
\nonumber \\
&=&2\sum_{q=1}^\infty
(-1)^q\left(\frac{1}{q}-
\frac{q \cos\frac{\pi}{N}}{q^2-1/N^2}-\frac{i}{N}\frac{\sin\frac{\pi}{N}
}{q^2-1/N^2}\right)
\nonumber \\
& &-i\pi+iN\sin\frac{\pi}{N}+N\lim_{\epsilon\rightarrow 0}
\frac{1}{\epsilon}
\label{ri}
\end{eqnarray}
Using the summation formula \cite{PrBrMa1},
\begin{equation}
2a^2\sum_{q=1}^\infty
\frac{(-1)^q}{q^2-a^2}=1-\frac{\pi a}{\sin\pi a}\,,
\end{equation}
we see the finite imaginary terms in (\ref{ri}) are canceled.
The last term in the right-hand side of (\ref{ri}) is divergent but
has no physical significance. We consider that the term should be dropped
as a consequence of regularization.%
\footnote{Namely, the regularization is consequently dropping the
contribution from $I_0$ in the series and taking only the real part of
the free energy.}

Now, the regularized free energy derived from the deformed heat kernel
trace of the conformal scalar field on $S^3$ is given as%
\footnote{Another
expression is $\bar{F}_3=\frac{1}{2\left(1-\cos\frac{2\pi}{N}\right)}
\left\{\ln2+\frac{1}{2}\cos\frac{\pi}{N}[\beta(1/N)+\beta(-1/N)]\right\}$,
$\beta(z)\equiv\frac{1}{2}[\psi((z+1)/2)-\psi(z/2)]$, where $\psi(z)$ is
the digamma function \cite{PrBrMa2}.}
\begin{eqnarray}
\bar{F}_3&=&\frac{1}{2\left(1-\cos\frac{2\pi}{N}\right)}\sum_{q=1}^\infty
(-1)^{q-1}\left(\frac{1}{q}-
\frac{q \cos\frac{\pi}{N}}{q^2-1/N^2}\right)
\nonumber \\
&=&\frac{1}{2\left(1-\cos\frac{2\pi}{N}\right)}\left[\ln2-
\cos\frac{\pi}{N}\left(\ln 2+\sum_{k=1}^\infty
\frac{\eta(2k+1)}{N^{2k}}\right)\right]
\nonumber \\
&=&\frac{\sin^2\frac{\pi}{2N}}{2\sin^2\frac{\pi}{N}}\ln2-
\frac{\cos\frac{\pi}{N}}{4\sin^2\frac{\pi}{N}}\sum_{k=1}^\infty
\frac{\eta(2k+1)}{N^{2k}}\nonumber \\
&=&\frac{1}{8\cos^2\frac{\pi}{2N}}\ln2-
\frac{\cos\frac{\pi}{N}}{4\sin^2\frac{\pi}{N}}\sum_{k=1}^\infty
\frac{\eta(2k+1)}{N^{2k}}\nonumber \\
&=&\frac{\ln 2}{8}-\frac{\eta(3)}{4\pi^2}+\frac{\pi^2}{N^2}
\left(\frac{\ln
2}{32}+\frac{\eta(3)}{24\pi^2}-\frac{\eta(5)}{4\pi^2}\right)
+O\left(\frac{\pi^4}{N^4}\right)\nonumber \\
&\approx&0.0638+0.0230\frac{\pi^2}{N^2}+O\left(\frac{\pi^4}{N^4}\right)\,.
\end{eqnarray}

The free energy from the deformed heat kernel trace of
the pseudo-conformal scalar field on $S^3$ is obtained
by setting $x=0$ in the formulas used above.
We find
\begin{eqnarray}
\bar{F}_{3pc}&=&\frac{1}{2\left(1-\cos\frac{2\pi}{N}\right)}\sum_{q=1}^\infty
\left(
\frac{q}{q^2-1/N^2}-\frac{1}{q}\right)
\nonumber \\
&=&\frac{1}{2N^2\left(1-\cos\frac{2\pi}{N}\right)}\sum_{q=1}^\infty
\frac{1}{q\left(q^2-1/N^2\right)}
\nonumber
\\&=&\frac{1}{4\sin^2\frac{\pi}{N}}\sum_{k=1}^\infty
\frac{\zeta(2k+1)}{N^{2k}}\nonumber
\\&=&\frac{\zeta(3)}{4\pi^2}+\frac{\pi^2}{N^2}
\left(\frac{\zeta(3)}{12\pi^{2}}+\frac{\zeta(5)}{4\pi^{4}}\right)
+O\left(\frac{\pi^4}{N^4}\right)\nonumber \\
&\approx&0.0304+0.0128\frac{\pi^2}{N^2}+O\left(\frac{\pi^4}{N^4}\right)\,.
\end{eqnarray}

\begin{figure}[ht]
\centering
\includegraphics[height=3.5cm]
{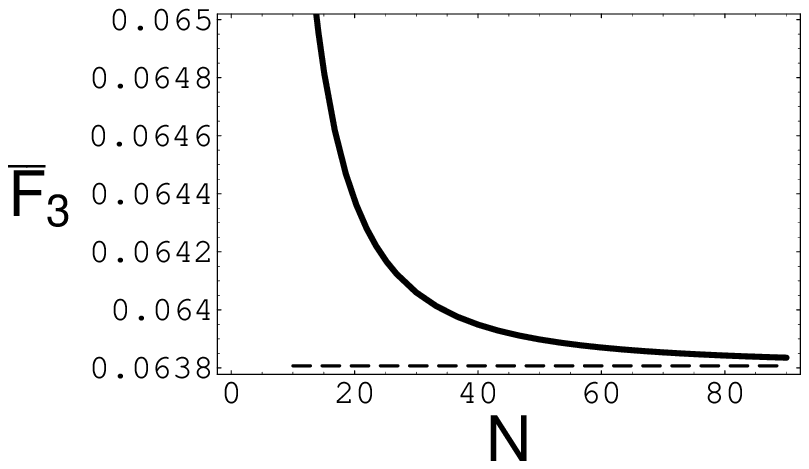}
\hspace{1cm}
\includegraphics[height=3.5cm]
{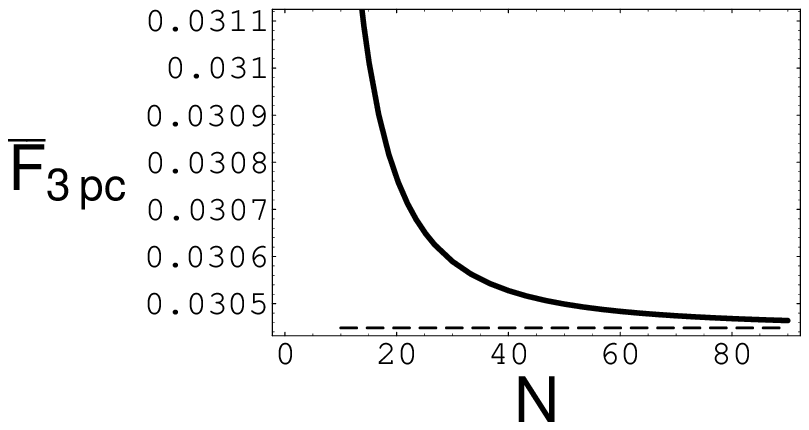}
\\
\hspace{1cm}(a)\hspace{6.5cm}(b)
\caption{%
(a) $\bar{F}_3$ is plotted against $N$. The horizontal dashed line
indicates $F_3$.
(b) $\bar{F}_{3pc}$ is plotted against $N$. The horizontal dashed line
indicates $F_{3pc}$. }
\label{fig2}
\end{figure}

The $N$-dependence of $\bar{F}_3$ and $\bar{F}_{3pc}$ is plotted in
FIG.~\ref{fig2}. The limit of $N\rightarrow\infty$ yields $F_3$ and
$F_{3pc}$, respectively. The difference from the limiting value is one
percent if
$N\approx 20$ in each case.

For brevity, we hereafter call the free energy from the deformed heat
kernel trace the `deformed free energy'.
In a similar manner, the deformed free energy of a conformal scalar field
$\bar{F}_5$ can be calculated. Using the summation formula, we find
\begin{eqnarray}
& &\int_0^\infty\frac{dt}{t}\sum_{p=0}^{N-1}(1-\cos\theta_p)
\left(\cos\theta_1-\cos\theta_p\right)\,
e^{2t\cos\theta_p-2t\cosh x}
\nonumber \\
&=&N
\int_0^\infty\frac{dt}{t}\sum_{q\in\mathbb{Z}}\left\{-
\frac{1}{2}\left(1+\cos\frac{2\pi}{N}\right)
\left[I_{Nq-1}(2t)-2I_{Nq}(2t)+I_{Nq+1}(2t)\right]\right.
\nonumber \\
&
&\left.\qquad\qquad\qquad+\frac{1}{4}\left[I_{Nq-2}(2t)-2I_{Nq}(2t)+I_{Nq+2}(2t)\right]\right\}
e^{-2t\cosh x}\,,
\end{eqnarray}
and we finally obtain
\begin{eqnarray}
\bar{F}_5&=&-\frac{1}{4\sin\frac{\pi}{N}\sin^2\frac{2\pi}{N}
\sin\frac{3\pi}{N}}\left[2\sin^3\frac{\pi}{2N}\sin\frac{3\pi}{2N}\ln
2\right.\nonumber \\
&
&\quad\quad\left.+\sum_{k=1}^\infty\left\{\cos^3\frac{\pi}{N}-2^{2(k-1)}
\cos\frac{2\pi}{N}
\right\}\frac{\eta(2k+1)}{N^{2k}}\right]\nonumber \\
&=&-\frac{1}{2^8}\left(2\ln
2+\frac{8\eta(3)}{3\pi^2}-\frac{16\eta(5)}{\pi^4}\right)\nonumber \\
& &-\frac{1}{2^8}\frac{\pi^2}{N^2}\left(5\ln
2+\frac{82\eta(3)}{9\pi^2}-\frac{40\eta(5)}{3\pi^4}
-\frac{80\eta(7)}{3\pi^6}\right)+O\left(\frac{\pi^4}{N^4}\right)
\nonumber \\
&\approx&-0.00574-0.0159\frac{\pi^2}{N^2}+O\left(\frac{\pi^4}{N^4}\right)\,.
\end{eqnarray}
In the limit of $N\rightarrow\infty$, one can find $\bar{F}_5\rightarrow
F_5=-\frac{1}{2^8}\left(2\ln
2+\frac{2\zeta(3)}{\pi^2}-\frac{15\zeta(5)}{\pi^4}\right)$
\cite{KlPuSa}.

The deformed free energy of a pseudo-conformal scalar field
$\bar{F}_{5pc}$ is derived as
\begin{eqnarray}
\bar{F}_{5pc}&=&-\frac{1}{4\sin\frac{\pi}{N}\sin^2\frac{2\pi}{N}
\sin\frac{3\pi}{N}}\sum_{k=1}^\infty\left\{2^{2(k-1)}-\cos^2\frac{\pi}{N}
\right\}\frac{\zeta(2k+1)}{N^{2k}}\nonumber \\
&=&-\frac{\zeta(3)}{48\pi^2}-\frac{\zeta(5)}{16\pi^4}
-\frac{\pi^2}{N^2}\left(\frac{\zeta(3)}{18\pi^2}+\frac{5\zeta(5)}{24\pi^4}+
\frac{5\zeta(7)}{16\pi^6}\right)+O\left(\frac{\pi^4}{N^4}\right)\nonumber
\\
&\approx&-0.00320-0.00931\frac{\pi^2}{N^2}+O\left(\frac{\pi^4}{N^4}\right)\,.
\end{eqnarray}
In the limit of $N\rightarrow\infty$, one can find
$\bar{F}_{5pc}\rightarrow
F_{5pc}=-\frac{\zeta(3)}{48\pi^2}-\frac{\zeta(5)}{16\pi^4}$.

\begin{figure}[ht]
\centering
\includegraphics[height=3.5cm]
{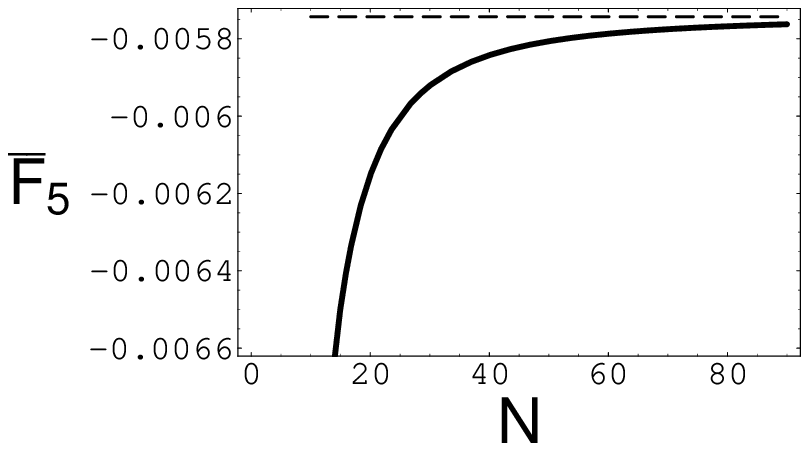}
\hspace{1cm}
\includegraphics[height=3.5cm]
{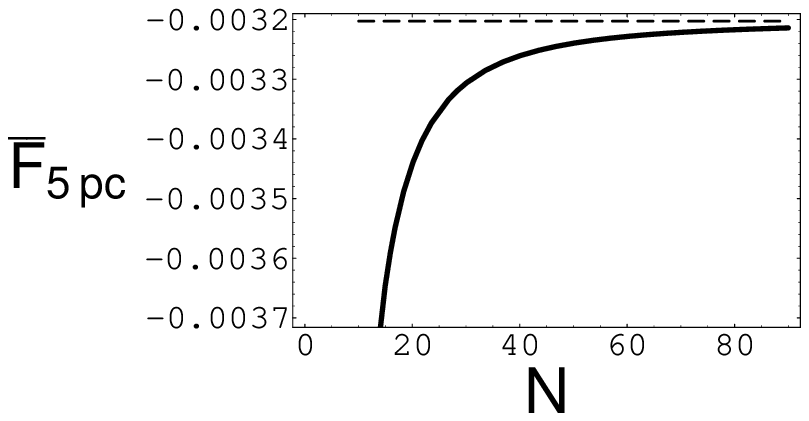}
\\
\hspace{1cm}(a)\hspace{6.5cm}(b)
\caption{%
(a) $\bar{F}_5$ is plotted against $N$. The horizontal dashed line
indicates $F_5$.
(b) $\bar{F}_{5pc}$ is plotted against $N$. The horizontal dashed line
indicates $F_{5pc}$. }
\label{fig3}
\end{figure}

The $N$-dependence of $\bar{F}_5$ and $\bar{F}_{5pc}$ is plotted in
FIG.~\ref{fig3}. The limit of $N\rightarrow\infty$ yields $F_5$ and
$F_{5pc}$, respectively. The difference from the limiting value is one
percent if
$N\approx 40$ in each case.

For a general odd dimension $d$, we find that the regularized expressions
of the deformed free energies take the form:%
\footnote{To obtain the integral forms, we utilize the identity
$\nu^{-1}=\int_0^\infty e^{-\nu\tau} d\tau$. For numerical calculations,
the convergence property of numerical integrations is rather worse than
that of the expressions in series sums.}
\begin{eqnarray}
\bar{F}_{d}&=&-\frac{1}{V_v(\cos\theta_v)}\sum_{n=0}^v\frac{b_{vn}}{2^n}
\sum_{r=0}^{\lfloor\frac{n-1}{2}\rfloor}\left(
\begin{array}{c}n \\ r\end{array}\right)\sum_{q=1}(-1)^q\left(
\frac{q \cos\frac{\pi
(n-2r)}{N}}{q^2-\frac{(n-2r)^2}{N^2}}-\frac{1}{q}\right)
\nonumber \\
&=&\frac{1}{2V_v(\cos\theta_v)}\sum_{n=1}^v {b_{vn}}\,\textrm{Re}\left[
\int_0^\infty\frac{N}{e^{N\tau}+1}\cosh^n\left(\tau-i\frac{\pi}{N}\right)\,
d\tau\right]
\nonumber \\
&=&\frac{1}{2V_v(\cos\theta_v)}
\int_0^\infty\frac{1}{e^{\tau}+1}\,\textrm{Re}\left[
V_v\left(\cosh\frac{\tau-i\pi}{N}\right)\right]\,
d\tau
\,,
\label{intfc}
\end{eqnarray}
and
\begin{eqnarray}
\bar{F}_{dpc}&=&-\frac{1}{N^2V_v(\cos\theta_v)}\sum_{n=1}^v\frac{b_{vn}}{2^n}
\sum_{r=0}^{\lfloor\frac{n-1}{2}\rfloor}\left(
\begin{array}{c}n \\ r\end{array}\right)\sum_{q=1}
\frac{(n-2r)^2}{q\left(q^2-\frac{(n-2r)^2}{N^2}\right)}
\nonumber \\
&=&\frac{1}{2V_v(\cos\theta_v)}\sum_{n=1}^v n{b_{vn}}
\int_0^\infty\ln(1-e^{-N\tau})\cosh^n\tau\,\tanh\tau\, d\tau
\nonumber \\
&=&\frac{1}{2V_v(\cos\theta_v)}
\int_0^\infty\ln(1-e^{-N\tau})V_v'(\cosh\tau)\,\sinh\tau\, d\tau
\nonumber \\
&=&-\frac{1}{2V_v(\cos\theta_v)}
\int_0^\infty\frac{N}{e^{N\tau}-1}V_v(\cosh\tau)\, d\tau
\nonumber \\
&=&-\frac{1}{2V_v(\cos\theta_v)}
\int_0^\infty\frac{1}{e^{\tau}-1}V_v\left(\cosh\frac{\tau}{N}\right)\,
d\tau
\,,
\label{intfpc}
\end{eqnarray}
where
\begin{equation}
V_v(\chi)=\prod_{n=0}^{v-1}\left(\cos\frac{2\pi n}{N}-\chi\right)\,,\quad
V_v'(\chi)=\frac{dV_v(\chi)}{d\chi}\,,
\end{equation}
and
\begin{equation}
b_{vn}=\frac{1}{n!}\left.\frac{d^n}{d\chi^n}V_v(\chi)\right|_{\chi=0}\,,
\end{equation}
i.e.,
\begin{equation}
V_v(\chi)=\sum_{n=0}^{v}b_{vn}\chi^n\,.
\end{equation}
Note that $V_v(1)=\sum_{n=0}^v b_n=0$.

The large $N$ limit of the integral representation yields
the normal free energy $F_d$ ($F_{dpc}$) on $S^d$ (with $d$ odd).
This is shown in Appendix B.

\section{A comment about $q$-analogue}
\label{qcom}

The normal derivation of degeneracies of Laplacian eigenvalues on $S^d$
is based on the following expression in terms of
the binomial coefficients:
\begin{equation}
g_l=\left(\begin{array}{c}d+l\\d\end{array}\right)-
\left(\begin{array}{c}d+l-2\\d\end{array}\right)=
\frac{(d+l)!}{d!l!}-\frac{(d+l-2)!}{d!(l-2)!}\,.
\end{equation}
In the literature of $q$-analysis \cite{KaCh},
the $q$-binomial coefficients are widely used.
Thus, a natural deformation of the degeneracy is given by
\begin{equation}
\bar{g}_p=\left[\begin{array}{c}d+p\\d\end{array}\right]-
\left[\begin{array}{c}d+p-2\\d\end{array}\right]
=\frac{[d+p]!}{[d]![p]!}-\frac{[d+p-2]!}{[d]![p-2]!}\,,
\end{equation}
where $[n]!=[n][n-1]\cdots[2][1]$.
The use of $q=e^{i\frac{2\pi}{N}}$ in this deformation yields
\begin{equation}
\hat{g}_p=\frac{\cos\frac{\pi}{N}(l+v)}{\cos\frac{\pi}{N}v}\bar{g}_p\,.
\end{equation}
The ambiguity in deformations comes from the fact that $[2 x]\ne [2][x]\ne
2[x]$.

Note that $\hat{g}_p$ can become negative.
To obtain the heat kernel trace, we may restrict $0\le p\le \lfloor
N/2\rfloor$, instead of dividing the trace by two.
If we want to use permutation invariance to take a simple trace,
we should consider the absolute value $|\hat{g}_p|$.
Using the expansion
\begin{equation}
\left|\cos\frac{\pi p}{N}\right|=\sqrt{1-\sin^2\frac{\pi
p}{N}}=1-\frac{1}{2}\sin^2\frac{\pi
p}{N}-\cdots\,,
\end{equation}
we can evaluate the $q$-deformed free energy using the degeneracy
$|\hat{g}_p|$, up to $O(N^{-2})$.
For example,
\begin{eqnarray}
\hat{F}_3&=&\frac{\ln 2}{8}-\frac{\eta(3)}{4\pi^2}+\frac{\pi^2}{N^2}
\left(\frac{5\ln
2}{64}-\frac{13\eta(3)}{48\pi^2}-\frac{5\eta(5)}{8\pi^2}\right)
+O\left(\frac{\pi^4}{N^4}\right)\nonumber \\
&\approx&0.0638+0.0232\frac{\pi^2}{N^2}+O\left(\frac{\pi^4}{N^4}\right)\,.
\end{eqnarray}
and
\begin{eqnarray}
\hat{F}_{3pc}&=&\frac{\zeta(3)}{4\pi^2}+\frac{\pi^2}{N^2}
\left(\frac{5\zeta(3)}{24\pi^2}+\frac{5\zeta(5)}{8\pi^2}\right)
+O\left(\frac{\pi^4}{N^4}\right)\nonumber \\
&\approx&0.0304+0.0320\frac{\pi^2}{N^2}+O\left(\frac{\pi^4}{N^4}\right)\,.
\end{eqnarray}
The deviation  at finite $N$ from the nondeformed free energy becomes even
slightly larger  than that in the case of our deformed free energy shown
in the previous section.

\section{Conclusion and prospect}
\label{conc}

In the present paper, we first computed the free energy of a massive
scalar field on a cycle graph $C_N$ by the heat kernel trace technique.
Motivated by the result, we calculated the trigonometric deformed version
of the free energy of a single conformally coupled free scalar field and
that of a single pseudo-conformally coupled free scalar field
on odd spheres.
We showed that the values of the deformed free energies, obtained through
the suitable regularization, approach the canonical values in the
limit $N\rightarrow\infty$.

As an analogue calculation, the deformation of the heat kernel trace
has many possible linkages to known theoretical models.

The calculations in the fuzzy space models \cite{Sa1,Sa2} and
the quantum geometry models \cite{CaOrTh2}
have a common feature that 
the heat kernel trace consists of the finite number of eigenvalues in the
models, though the degeneracy in such models is still integer.
The models with $q$-deformed extra dimensions \cite{NaTo} and
the statistical models with $q$-deformed algebra \cite{LaNa}
may lead to similar forms of heat kernel traces.

Although we have not yet convinced of the precise relation to our
calculation with noninteger degeneracies,
we suppose that some models such as the theory with deformed phase
spaces \cite{SVW,AGMA}, the nonlinear field space theory \cite{MiTr},
field theories on fractals \cite{DFractal},
and the unparticle models \cite{Ge,St,Le,NiSp,FrNiPa} might induce novel
spectral densities and 
nonstandard heat kernel traces. 
The study on their appropriate regularization is very interesting. Seeking
similarities to the heat kernels considered in random walks and quantum
walks
\cite{Kempe} is another subject of much interest.

It is interesting to imagine
inverse problems on heat kernel traces.
For example, we can ask whether there is an interacting scalar field
theory on $S^d$ having the similar heat kernel trace, which approaches
free theory if a certain number $N$ becomes large.
We wish to discover the physical significance of the
mathematical extension in such a sense. 

The generalization of the deformation to other functions and manipulations
is considered to be interesting.
There are some $q$-analogues of the exponential function
\begin{equation}
\sum_{n=0}^\infty \frac{x^n}{[n]!}\,,\quad 
\sum_{n=0}^\infty \frac{q^{\frac{n(n+1)}{2}}x^n}{[n]!}\,,
\end{equation}
and the function used in the nonextensive statistical mechanics
\cite{Tsallis}, in addition.
Moreover, the standard integration over the parameter $t$ is even able to
be changed along with the known extension in $q$-calculus \cite{KaCh}.
These general treatments, however, would go beyond the scope of the
present work.



\appendix

\section*{Appendix A}

In this Appendix A we present a direct calculation of the free energy of a
pseudo-conformal scalar field on odd spheres.

First, we regularize the free energy by discarding $l=0$ in the summation
in (\ref{smf}). For example of $d=3$, we take
\begin{equation}
F_{3pc}=-\frac{1}{2}\lim_{s\rightarrow 0}\sum_{l=1}^\infty
l^2\int_0^\infty\frac{dt}{t^{1-s}}e^{-l^2t}\,.
\end{equation}
Next, we interpret the integration over $t$ here as `$\Gamma(0)$',
and proceed as follows:
\begin{eqnarray}
F_{3pc}&=&-\frac{1}{2}\lim_{s\rightarrow 0}\Gamma(s)\sum_{l=1}^\infty
l^{2-2s}
=-\frac{1}{2}\lim_{s\rightarrow 0}\Gamma(s)\zeta(-2+2s)
\nonumber \\
&=&-\frac{1}{2}\lim_{s\rightarrow
0}(-1+s)\Gamma(-1+s)\zeta(-2+2s)\nonumber
\\ &=&\frac{1}{2}\pi^{-5/2}\lim_{s\rightarrow
0}\Gamma\left(\frac{3}{2}-s\right)\zeta(3-2s) =\frac{\zeta(3)}{4\pi^2}\,, 
\end{eqnarray}
where we have used the formula \cite{GrRy}
\begin{equation}
\Gamma\left(\frac{z}{2}\right)\zeta(z)=\pi^{z-\frac{1}{2}}
\Gamma\left(\frac{1-z}{2}\right)\zeta(1-z)\,.
\end{equation}
Now, we verify the previous result.

Similarly, a slight calculation shows the value of $F_{5pc}$ as follows:
\begin{eqnarray}
&&F_{5pc}=-\frac{1}{2}\lim_{s\rightarrow
0}\Gamma(s)\sum_{l=1}^\infty\frac{l^{2-2s}(l^2-1)}{12}
=-\frac{1}{24}\lim_{s\rightarrow
0}\Gamma(s)[\zeta(-4+2s)-\zeta(-2+2s)]\nonumber \\
&&=
-\frac{1}{24}\lim_{s\rightarrow
0}[(-1+s)(-2+s)\Gamma(-2+s)\zeta(-4+2s)-(-1+s)\Gamma(-1+s)\zeta(-2+2s)]\nonumber
\\
&&=-\frac{1}{24}\left[\frac{2}{\pi^{9/2}}\Gamma\left(\frac{5}{2}\right)
\zeta(5)+\frac{1}{\pi^{5/2}}\Gamma\left(\frac{3}{2}\right)\zeta(3)\right]
=-\frac{\zeta(3)}{48\pi^2}-\frac{\zeta(5)}{16\pi^4}\,.
\end{eqnarray}

The results of calculations of $F_{dpc}$ with those of $F_d$ obtained by
Klebanov, Pufu and Safdi \cite{KlPuSa} are shown in Table~\ref{table1}.

\begin{table}[htb]
\begin{minipage}[t]{\textwidth}
\begin{center}
{
\begin{tabular}{c|l}
$d$& $F_d$  (Ref.~\cite{KlPuSa}) \\ \hline
$3$ &$\frac{1}{2^4}\left(2\ln 2-\frac{3\zeta(3)}{\pi^2}\right)\approx
0.0638$\\
$5$ &$\frac{-1}{2^8}\left(2\ln 2+\frac{2\zeta(3)}{\pi^2}-
\frac{15\zeta(5)}{\pi^4}\right)\approx -5.74\times 10^{-3}$\\
$7$ &$\frac{1}{2^{12}}\left(4\ln 2+\frac{82\zeta(3)}{15\pi^2}-
\frac{10\zeta(5)}{\pi^4}-\frac{63\zeta(7)}{\pi^6}\right)\approx
7.97\times 10^{-4}$\\ 
$9$ &$\frac{-1}{2^{16}}\left(10\ln 2+\frac{1588\zeta(3)}{105\pi^2}-
\frac{2\zeta(5)}{\pi^4}-\frac{126\zeta(7)}{\pi^6}-
\frac{255\zeta(9)}{\pi^8}\right)\approx -1.31\times 10^{-4}$\\ 
$11$ &$\frac{1}{2^{20}}\left(28\ln 2+\frac{7794\zeta(3)}{175\pi^2}+
\frac{1940\zeta(5)}{63\pi^4}-\frac{1218\zeta(7)}{5\pi^6}-
\frac{850\zeta(9)}{\pi^8}-\frac{1023\zeta(11)}{\pi^{10}}\right)\approx
2.37\times 10^{-5}$\\
 & \hspace{13.5cm} \\
\end{tabular}
}
\end{center}
\end{minipage}
\\
\vspace{5mm}
\\
\begin{minipage}[t]{\textwidth}
\begin{center}
{
\begin{tabular}{c|l}
$d$& $F_{dpc}$ \\ \hline
$3$ &
$\frac{\zeta(3)}{4\pi^2}\approx 0.0304$\\ 
$5$ &
$-\frac{\zeta(3)}{48\pi^2}-\frac{\zeta(5)}{16\pi^2}\approx -3.20\times
10^{-3}$\\ 
$7$ &
$\frac{\zeta(3)}{360\pi^2}+\frac{\zeta(5)}{96\pi^4}+
\frac{\zeta(7)}{64\pi^6}\approx 4.66\times 10^{-4}$\\ 
$9$ &
$-\frac{\zeta(3)}{2240\pi^2}-\frac{7\zeta(5)}{3840\pi^4}-
\frac{\zeta(7)}{256\pi^6}-\frac{\zeta(9)}{256\pi^8}\approx -7.83\times
10^{-5}$\\ 
$11$ &
$\frac{\zeta(3)}{12600\pi^2}+\frac{41\zeta(5)}{120960\pi^4}+
\frac{13\zeta(7)}{15360\pi^6}+\frac{\zeta(9)}{768\pi^8}+
\frac{\zeta(11)}{1024\pi^{10}}\approx 1.43\times 10^{-5}$\\  
 & \hspace{13.5cm} \\
\end{tabular}
}
\end{center}
\end{minipage}
\caption{The values of $F_d$ \cite{KlPuSa} and $F_{dpc}$.
}
\label{table1}
\end{table}

The integral expression of $F_d$ has been known as \cite{GiKl}
\begin{equation}
\tilde{F}_d\equiv-\sin\frac{\pi d}{2}F_d=\frac{1}{\Gamma(1+d)}
\int_0^1du\,u\,\sin\pi u\, \Gamma\left(\frac{d}{2}+u\right)
\Gamma\left(\frac{d}{2}-u\right)\,.
\label{intc}
\end{equation}
We happened to find the following similar expression of $F_{dpc}$:
\begin{equation}
\tilde{F}_{dpc}\equiv-\sin\frac{\pi d}{2}F_{dpc}=\frac{1}{\Gamma(1+d)}
\int_{-1/2}^{1/2}du\,u\,\sin\pi u\, \Gamma\left(\frac{d}{2}+u\right)
\Gamma\left(\frac{d}{2}-u\right)\,.
\label{intpc}
\end{equation}
The derivation of these integral expressions is performed in Appendix B.

\section*{Appendix B}

In this Appendix B, we show that the integral representations in the last
line of (\ref{intfc}) and (\ref{intfpc}) in Section~\ref{calcdef} can be
utilized to obtain the values of
$F_d$ and
$F_{dpc}$ respectively by taking the limit of $N\rightarrow\infty$.

To this end, we first observe 
\begin{equation}
V_v(\cos\theta_v)\rightarrow\frac{2^v\pi^{2v}}{N^{2v}}\prod_{n=0}^{v-1}
(v^2-n^2)=\frac{2^{v-1}\pi^{2v}}{N^{2v}}(2v)!\,,
\end{equation}
and
\begin{equation}
V_v\left(\cosh\frac{w}{N}\right)\rightarrow\frac{2^v\pi^{2v}}{N^{2v}}\prod_{n=0}^{v-1}
\left[-n^2-\left(\frac{w}{2\pi}\right)^2\right]
=\frac{2^v\pi^{2v}}{N^{2v}}
\frac{\Gamma\left(v-i\frac{w}{2\pi}\right)
\Gamma\left(v+i\frac{w}{2\pi}\right)}{\Gamma\left(-i\frac{w}{2\pi}\right)
\Gamma\left(i\frac{w}{2\pi}\right)}\,,
\end{equation}
at large $N$.
Then, we get
\begin{eqnarray}
&
&\lim_{N\rightarrow\infty}\bar{F}_d=F_d\nonumber \\
& &=-\frac{(-1)^v}{(2v)!}\textrm{Re}\left[\int_0^\infty
\Gamma\left(v-i\left(\tau-{i}/{2}\right)\right)
\Gamma\left(v+i\left(\tau-{i}/{2}\right)\right)
\left(\tau-{i}/{2}\right)e^{-\pi\left(\tau-\frac{i}{2}\right)}
 d\tau
\right]\,,
\label{intic}
\end{eqnarray}
and
\begin{equation}
\lim_{N\rightarrow\infty}\bar{F}_{dpc}=F_{dpc}
=-\frac{(-1)^v}{(2v)!}\int_0^\infty
\Gamma\left(v-i\tau\right)
\Gamma\left(v+i\tau\right)t e^{-\pi\tau}
\,d\tau\,.
\label{intipc}
\end{equation}
To obtain these expressions, we have employed the identities \cite{GrRy}
\begin{equation}
\Gamma(1-z)\Gamma(z)=\frac{\pi}{\sin\pi z}\,,\quad
\Gamma(z+1)=z\Gamma(z)\,.
\end{equation}

Their relations to the other expressions (\ref{intc}) and (\ref{intpc}) in
Appendix A can be found by considering integrals in complex domain.

\begin{figure}[ht]
\centering
\hspace{1cm}
\includegraphics[height=4cm]
{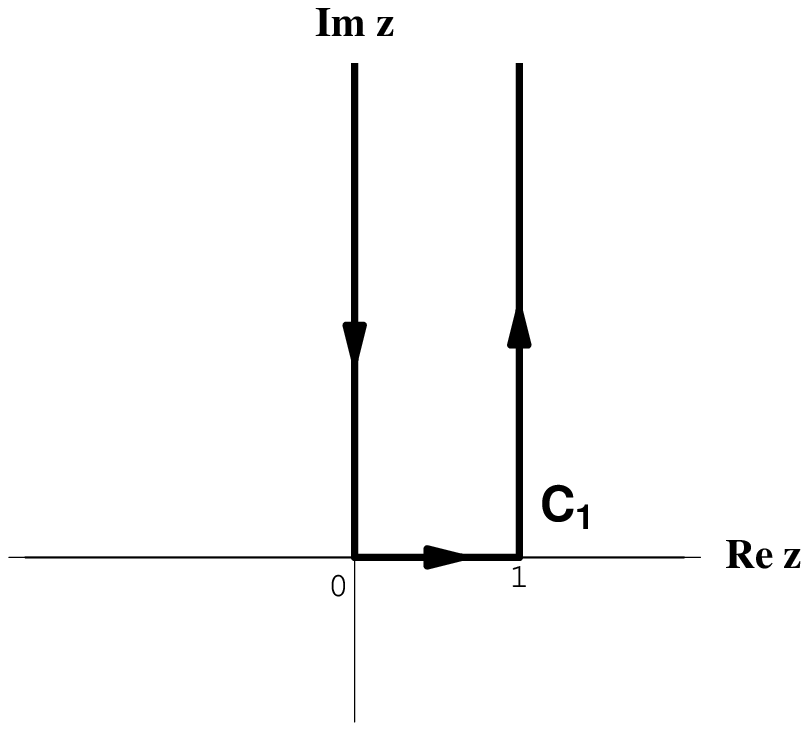}
\hspace{1cm}
\includegraphics[height=4cm]
{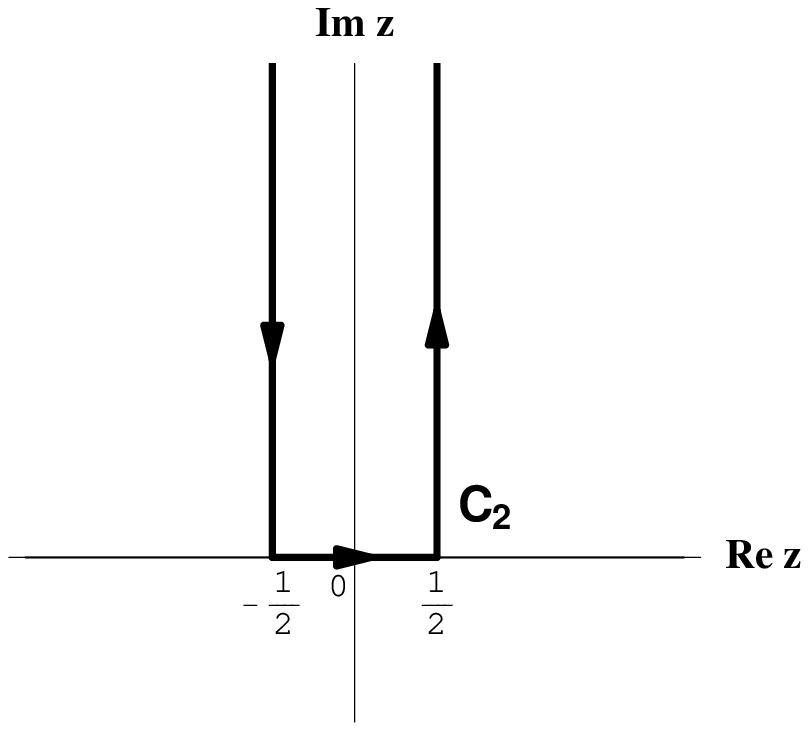}
\\
\hspace{0.5cm}
(a)\hspace{5cm}(b)
\caption{%
Contours $\textrm{C}_1$ (a) and $\textrm{C}_2$ (b).}
\label{contc}
\end{figure}
To find the relation on $F_d$, consider the integral
\begin{equation}
\int_{\textrm{C}_1}
\Gamma\left(v+\frac{1}{2}-z\right)\Gamma\left(v+\frac{1}{2}+z\right)
z e^{i\pi z} dz\,,
\end{equation}
with $\textrm{C}_1$ the contour of FIG.~\ref{contc}(a).
Because of the absence of poles inside the contour,
the integral vanishes. The fact leads to
\begin{eqnarray}
& &-\int_0^\infty
\Gamma\left(v+\frac{1}{2}-i\tau\right)\Gamma\left(v+\frac{1}{2}+i\tau\right)
(i\tau) e^{-\pi\tau} i d\tau\nonumber \\
& &+\int_0^1
\Gamma\left(v+\frac{1}{2}-u\right)\Gamma\left(v+\frac{1}{2}+u\right)
u e^{i\pi u} du\nonumber \\
& &+\int_0^\infty
\Gamma\left(v+\frac{1}{2}-1-i\tau\right)
\Gamma\left(v+\frac{1}{2}+1+i\tau\right)
(i\tau+1) (-1) e^{-\pi\tau} i d\tau=0\,.
\end{eqnarray}
Arranging the equality, we obtain
\begin{eqnarray}
& &\int_0^1
\Gamma\left(v+\frac{1}{2}-u\right)\Gamma\left(v+\frac{1}{2}+u\right)
u e^{i\pi u} du\nonumber \\
& &=-\int_0^\infty
\tau\left(v-\frac{1}{2}-i\tau\right)\Gamma\left(v-\frac{1}{2}-i\tau\right)
\Gamma\left(v+\frac{1}{2}+i\tau\right)
e^{-\pi\tau}  d\tau\nonumber \\
& &-\int_0^\infty
(\tau-i)
\left(v+\frac{1}{2}+i\tau\right)\Gamma\left(v-\frac{1}{2}-i\tau\right)
\Gamma\left(v+\frac{1}{2}+i\tau\right)
 e^{-\pi t}  d\tau\nonumber \\
& &=-(2v+1)\int_0^\infty
\Gamma\left(v-\frac{1}{2}-i\tau\right)
\Gamma\left(v+\frac{1}{2}+i\tau\right)\left(\tau-\frac{i}{2}\right)
e^{-\pi\tau}  d\tau\,.
\label{rp}
\end{eqnarray}
Comparing the imaginary part of each side of (\ref{rp}), we find that
Eq.~(\ref{intc}) is equivalent to Eq.~(\ref{intic}).

To find the relation on $F_{dpc}$, consider the similar integral
\begin{equation}
\int_{\textrm{C}_2}
\Gamma\left(v+\frac{1}{2}-z\right)\Gamma\left(v+\frac{1}{2}+z\right)
z e^{i\pi z} dz=0\,,
\end{equation}
with $\textrm{C}_2$ the contour of FIG.~\ref{contc}(b).
This leads to
\begin{eqnarray}
& &-\int_0^\infty
\Gamma\left(v+\frac{1}{2}+\frac{1}{2}-i\tau\right)
\Gamma\left(v+\frac{1}{2}-\frac{1}{2}+i\tau\right)
\left(i\tau-\frac{1}{2}\right) (-i) e^{-\pi\tau} i d\tau\nonumber \\
& &+\int_{-1/2}^{1/2}
\Gamma\left(v+\frac{1}{2}-u\right)\Gamma\left(v+\frac{1}{2}+u\right)
u e^{i\pi u} du\nonumber \\
& &+\int_0^\infty
\Gamma\left(v+\frac{1}{2}-\frac{1}{2}-i\tau\right)
\Gamma\left(v+\frac{1}{2}+\frac{1}{2}+i\tau\right)
\left(i\tau+\frac{1}{2}\right) (i) e^{-\pi\tau} i d\tau=0\,.
\end{eqnarray}
Arranging the equality, we obtain
\begin{eqnarray}
& &\int_{-1/2}^{1/2}
\Gamma\left(v+\frac{1}{2}-u\right)\Gamma\left(v+\frac{1}{2}+u\right)
u e^{i\pi u} du\nonumber \\
& &=\int_0^\infty
\left(i\tau-\frac{1}{2}\right)
\left(v-i\tau\right)\Gamma\left(v-i\tau\right)
\Gamma\left(v+i\tau\right)
e^{-\pi\tau}  d\tau\nonumber \\
& &+\int_0^\infty
\left(i\tau+\frac{1}{2}\right)\left(v+i\tau\right)
\Gamma\left(v-i\tau\right)
\Gamma\left(v+i\tau\right)
 e^{-\pi\tau}  d\tau\nonumber \\
& &=i(2v+1)\int_0^\infty
\Gamma\left(v-i\tau\right)
\Gamma\left(v+i\tau\right)\tau
e^{-\pi\tau}  d\tau\,.
\label{rpc}
\end{eqnarray}
Comparing the imaginary part of each side of (\ref{rpc}), we find that
equations (\ref{intpc}) and (\ref{intipc}) are equivalent.



\end{document}